\documentclass[fleqn,10pt]{wlscirep}

\usepackage{bm}

\usepackage{color}

\title{Enhancement of the thermoelectric properties in bilayer graphene structures induced by Fano resonances}

\author[1]{J. A. Briones-Torres}
\author[1,*]{R. P\'{e}rez-\'{A}lvarez}
\author[2]{S. Molina-Valdovinos}
\author[2,**]{I. Rodr\'{i}guez-Vargas}
\affil[1]{Centro de Investigaci\'{o}n en Ciencias, Universidad Aut\'{o}noma del Estado de Morelos, Av. Universidad 1001 Col. Chamilpa, 62209 Cuernavaca, Morelos, M\'{e}xico.}
\affil[2]{Unidad Acad\'{e}mica de Ciencia y Tecnolog\'{i}a de la Luz y la Materia, Universidad Aut\'{o}noma de Zacatecas, Carretera Zacatecas-Guadalajara Km. 6, Ejido La Escondida, 98160 Zacatecas, Zac., Mexico.}

\affil[*]{rpa@uaem.mx}
\affil[**]{isaac@uaz.edu.mx}


\begin{abstract}

Fano and hybrid resonances of bilayer graphene could be attractive for thermoelectric devices. The special profile presented by such resonances could significantly enhance the Seebeck coefficient and the power factor. In this work, we study the thermoelectric properties of bilayer graphene single and double barrier structures. The charge carriers are described as massive chiral particles through an effective Dirac-like Hamiltonian. The Hybrid matrix method, the Landauer-B\"uttiker formalism and the Cutler-Mott formula are implemented to obtain the transmission, transport and thermoelectric properties, respectively. The Seebeck coefficient and the power factor are analyzed for gapless and gapped single and double barriers. We find that in the energy range where Fano resonances occur, the Seebeck coefficient attains values of tens of mV/K and the power factor reaches values of the order nW/K$^{2}$. Hybrid resonances also sustain high values for the thermoelectric properties, however not as high as Fano resonances. We also find that despite the Fano and hybrid profiles are manifested in the conductance of gapped barrier structures, the Seebeck coefficient and the power factor are systematically reduced as the bandgap gets larger. So, our findings indicate that bilayer graphene barrier structures can be used to improve the response of thermoelectric devices. 

\end{abstract}

\begin{document}

\flushbottom
\maketitle

\thispagestyle{empty}

\section*{Introduction}

At present, there is a strong tendency to use sustainable, renewable and clean energy to improve the quality of human life.\cite{1,2} In particular, thermoelectricity a phenomenon that involves heat energy has been a topic of interest.\cite{3,4,5} A thermoelectric device converts heat energy into electrical energy and vice versa.\cite{3,4} The process by which electricity is generated through a temperature gradient is known as the Seebeck effect.\cite{3,4,6,7} Under this context, there is a search for materials with outstanding thermoelectric properties.\cite{8,9} The parameter that helps us to identify a good thermoelectric material is the so-called Figure of Merit,\cite{7} given by $ZT = \frac{\sigma S^{2}T}{\kappa}$, where $\sigma$ is the electric conductivity, $S$ is the Seebeck coefficient, $\kappa$ the thermal conductivity and $T$ is the temperature. A high value of $ZT$ translates into high thermoelectric efficiency.\cite{8,9} However, for high $ZT$, we need high $S$ and $\sigma$ as well as small $\kappa$. This is difficult because the Wiedemann-Franz law \cite{10,11} and the Mott law \cite{11} impose certain restrictions on the ratio $\sigma/\kappa$ and on the product $\sigma S^{2}$, respectively. The latter is known as the power factor and is directly related to the efficiency of thermoelectric devices.\cite{3} This is where the search for materials that maximize the figure of merit comes into play. Generally in conventional semiconductors and mixtures between them, values of around 1 have been reached.\cite{12,13} We are talking about materials such as bismuth, tellurium, silicon and germanium. In the case of nanostructures, values around 1.5 have been obtained.\cite{12,13} Recent research in 2D materials such as graphene \cite{1,2,14} and silicene \cite{15,16,17} have reported values above 2 and even above the technological limit value $ZT = 3$.\cite{18,19,20,21,22} Working with nanostructured 2D materials has become a trend,\cite{18,20,21,22,23,24,25} not only for its excellent properties, but also for the possible redistribution of the density of states. A high accumulation of states is linked to high values of $ZT$.\cite{18,22,25} One of the materials that promises a lot to the technology industry is bilayer graphene.\cite{26,27,28} This material is quasi-two-dimensional, with excellent electrical conductivity,\cite{29} high thermal conductivity \cite{30} and a modulable bandgap with the application of an external electric field.\cite{14,31} Bilayer graphene also harbors exotic effects that could substantially improve the power factor, and consequently giving rise to high values of $ZT$. Among these effects we can find anti-Klein tunneling,\cite{33,34} cloaked states,\cite{34,35} and Fano and hybrid resonances.\cite{36,37,38,39,40}\\

In the case of Fano resonances, it has been shown that they come from the chiral nature of electrons in bilayer graphene.\cite{40} In particular, they arise due to the chiral matching between electron states inside and outside electrostatic barriers at oblique incidence.\cite{36,37,38,39,41} On its side, hybrid resonances are the result of the coupling between Fano resonances and resonant states in double and superlattice barrier structures.\cite{39,41} Both types of resonance can retain its special profile if the angle of incidence is restricted to values close to normal incidence.\cite{39} They can also leave a characteristic mark on transport properties that can help to corroborate this exotic phenomenon of Fano resonances in bilayer graphene.\cite{39} Moreover, Fano and hybrid profiles can be manifested in the conductance by opening and modulating a bandgap.\cite{Briones-Torres2020} Taking into account the special profile of Fano and hybrid resonances as well as its impact on the transport properties of bilayer graphene barrier structures, we can expect high values for the Seebeck coefficient and the power factor. Actually, there are several reports in the literature in which is well-documented how Fano resonances enhance thermoelectricity.\cite{finch2009,gomezsilva2012,garciasuarez2013,liu2014,wang2016,taniguchi2020} The typical systems consist of quantum-dot interferometers, molecular junctions and chain junctions to mention a few. The common factor in all these studies is the quantum control of Fano resonances to improve the thermoelectric properties. However, as far as we know,	 there are no reports dealing with the impact of Fano and hybrid resonances on the thermoelectric properties of bilayer graphene barrier structures. In fact, most works about thermoelectricity in bilayer graphene report the Seebeck coefficient in doped systems,\cite{42} systems with applied magnetic fields,\cite{43} disorder \cite{44} and bandgap opening.\cite{45} So, we consider that a thorough assessment of the impact of Fano and hybrid resonances on the thermoelectric properties of bilayer graphene barrier structures is necessary.\\

In the present work, we calculate the Seebeck coefficient and the power factor in bilayer graphene single and double barriers. The hybrid matrix method and the Landauer-B\"uttiker formalisms are used to calculate the transmittance and conductance, respectively.\cite{41,46,47,48} Likewise, the Seebeck coefficient and the power factor are obtained using the Cuttler-Mott formula.\cite{49,50} We analyze energy regions where Fano and hybrid resonances contribute significantly to the transport properties, finding that the Seebeck coefficient and the power factor reach values of mV/K and nW/K$^2$, respectively. 

\section*{Methodology}

%The system that we are interested in is a bilayer graphene superlattice. In particular, we will deal with BGSLs conceived by metallic electrodes arranged in periodic fashion, see Fig. \ref{fig:Fig1}a. Through the electrodes we can apply an electrostatic field perpendicularly to the graphene sheets in such way that the electrostatic potential in the two layers be the same. Keeping the same potential between the two layers ensures that the symmetry between them is preserved and consequently that no bandgap opening arises in the band structure. Then, the main effect of the electrostatic potential is a shifting of the Dirac paraboloids along the energy axis, see Fig. \ref{fig:Fig1}b. By arranging regions with and without electrostatic potential in periodic fashion we can generate a typical superlattice band-edge profile, see Fig. \ref{fig:Fig1}a. The transmission and transport properties of this system can be obtained through the hybrid matrix method \cite{33,34} and the Landauer-B\"uttiker formalism \cite{42}, respectively. The basic information that we need in order to implement the mentioned methodologies is related to the wave functions and dispersion relations in the two regions that define the unit cell of the superlattice, that is, the region with and without electrostatic potential. So, we will present in first place the wave functions and dispersion relations to proceed with the generals of the hybrid matrix method and the Landauer-B\"uttiker formalism. \\ 

\begin{figure}[htb]
\centering
\includegraphics[scale=0.75]{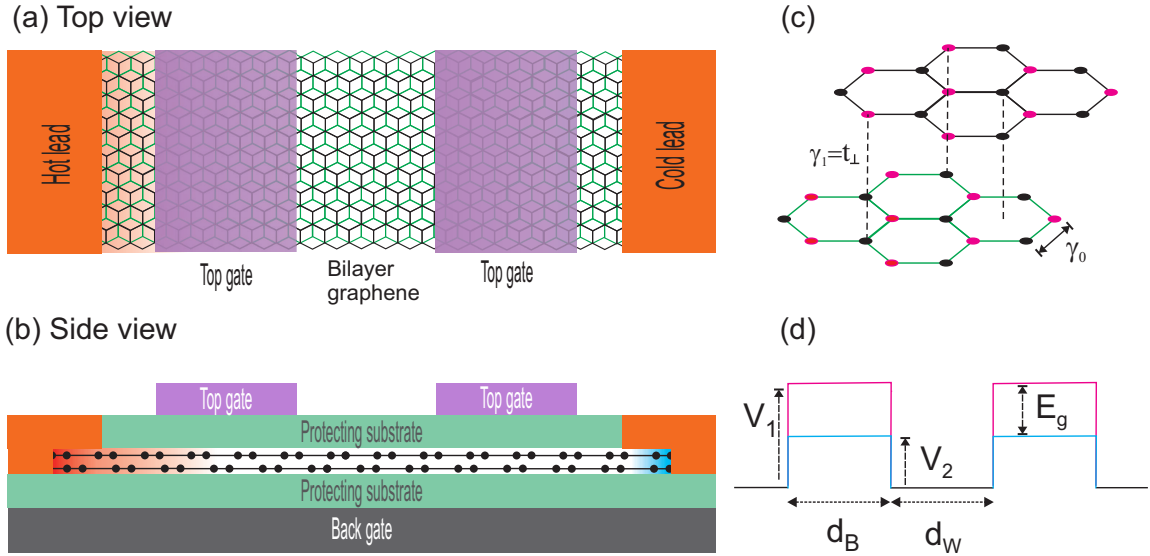}
\caption{\label{Fig0} (a) Top and (b) side view of the schematic representation of the possible thermoelectric device based on bilayer graphene double barriers. The double barrier structure generated by the top gates is sandwiched between hot and cold leads. (c) Crystal structure of Bernal-stacked bilayer graphene, with $\gamma_{0}$ and $\gamma_{1}= t_{\perp}$ as the main interactions between carbon atoms. (d) Band-edge profile of the conduction band of (a). Here, $V_{1}$ and $V_{2}$ represent the potential energy in the top and bottom graphene sheet, respectively. $E_{g}=V_{2}-V_{1}$ is the bandgap energy. $d_{B}$ and $d_{w}$ represents the widths of the barrier and well regions, respectively.}
\end{figure}

In figure  \ref{Fig0}a and b we illustrate the possible thermoelectric device based on bilayer graphene barrier structures. As a representative case we are showing a double-barrier thermoelectric device. The heat flows from the left to the right lead, hot and cold side, respectively. Bilayer graphene is placed on a substrate, which in turn is located above a back gate. The potential barriers are generated by electrodes (gates) on top of bilayer graphene. The crystal structure is schematically shown in figure \ref{Fig0}c. It corresponds to stable Bernal-stacked bilayer graphene. The relevant interactions between carbon atoms for our Hamiltonian are depicted as well, $\gamma_{0}$ and $\gamma_{1}=t_{\perp}$. The potential barriers are illustrated in figure \ref{Fig0}d. Gapless or gapped barriers can be obtained by applying the same or different potential energies to the graphene sheets. 

To study the thermoelectric properties of bilayer graphene structures we will use the stable numerical method known as the hybrid matrix method,\cite{41,46,47} within the context of the Sturm-Liouville formalism.\cite{41,47} To establish the hybrid matrix method, we need firstly the effective Dirac Hamiltonian that describes the charge carriers in bilayer graphene, namely:\cite{28,51,52}

\begin{equation}\label{eq1}
\mathcal{H}=\left(
\begin{array}{cccc}
V_{1}&\pi&t_{\perp}&0\\
\pi^{*}&V_{1}&0&0\\
t_{\perp}&0&V_{2}&\pi^{*}\\
0&0&\pi&V_{2}
\end{array}
\right),
\end{equation}

\noindent where $\pi=v_{F}(p_{x}+ip_{y})$, $\pi^{*}=v_{F}(p_{x}-ip_{y})$, $p_{x,y}=-i\hbar\partial_{x,y}$ is the momentum operator and $v_{F}$ is the Fermi velocity. This Hamiltonian is valid for energies less than $\gamma_{0}=3.09$ eV,\cite{51,52} it involves the interaction between graphene layers through the parameter $t_{\perp}=390$ meV \cite{52} and takes into account bandgap opening $E_{g}=V_{2}-V_{1}$ when $V_1 \neq V_2$.\\
\noindent Now, it is important to mention that in our single and double barrier structures there is spatial homogeneity in the transverse direction, that is $q_{y}=k_{y}$, then the propagation of Dirac electrons can be reduced to a one-dimensional problem. So, we need secondly to solve the one-dimensional eigenvalue equation $\mathbf{H}\cdot\mathbf{F}(x)=E\mathbf{F}(x)$, which after basic algebra adopts the mathematical form\cite{Briones-Torres2020}

\begin{equation}\label{eq2}
\frac{d\mathbf{F}(x)}{dx}+\left(
\begin{array}{cccc}
q_{y}&i\frac{V_{1}-E}{\hbar v_{F}}&0&0\\
i\frac{V_{1}-E}{\hbar v_{F}}&-q_{y}&i\frac{t_{\perp}}{\hbar v_{F}}&0\\
0&0&-q_{y}&i\frac{V_{2}-E}{\hbar v_{F}} \\
i\frac{t_{\perp}}{\hbar v_{F}}&0&i\frac{V_{2}-E}{\hbar v_{F}}&q_{y}
\end{array}\right)\cdot\mathbf{F}(x)=\mathbf{0}_{4\times1}.
\end{equation}

\noindent Some modifications have been made to the Sturm-Liouville formalism in order to use linear equations that do not involve second order derivatives. Then to find a basis, a function of the form $\mathbf{F}(x)=\mathbf{F}_{0}e^{iqx}$ is proposed, this is substituted in equation (\ref{eq2}) to arrive at a secular equation that has the following eigenvalues as a solution

\begin{equation}\label{eq3}
\small{
q=\pm\sqrt{-q^{2}_{y}-\frac{1}{2(\hbar v_{F})^{2}}[(E-V_{1})^{2}+(E-V_{2})^{2}]\pm\frac{1}{2(\hbar v_{F})^{2}}\sqrt{[(E-V_{1})^{2}-(E-V_{2})^{2}]^{2}+4t_{\perp}(E-V_{1})(E-V_{2})}}.}
\end{equation}

\noindent We can see that there are four eigenvalues, this comes from the fact that the Hamiltonian is 4$\times$4, then we have four linear independent solutions given as 

\begin{equation}\label{eq4}
\mathbf{F}(x)^{\pm}_{j}=\mathbf{F}_{0j}^{\pm}e^{iq_{j}x} = (a_{j},b_{j}^{\pm},c_{j},d_{j}^{\pm})^{T}e^{iq_{j}x},\quad $j=1,2$.
\end{equation}

\noindent where the corresponding components are 

\begin{eqnarray}\label{eq5}
a_{j} &=& i\frac{E-V_{1}}{\hbar v_{F}};\\
b_{j}^{\pm} &=& q_{y}\pm iq_{j};\\
c_{j} &=& \frac{i}{t_{\perp} \hbar v_{F}}[(E-V_{1})^{2}-(q_{y}^{2}+q_{j}^{2})(\hbar v_{F})^{2}];\\
d_{j}^{\pm} &=& \frac{t_{\perp}^{2} (E-V_{1})-(E-V_{2})[(E-V_{1})^{2}-(q_{y}^{2}+q_{j}^{2})(\hbar v_{F})^{2}]}{(q_{y}\pm iq_{j})t_\perp (\hbar v_{F})^{2}}. 
\end{eqnarray}

\noindent A general solution can be expressed in matrix form as follows

\begin{equation}\label{eq6}
\mathbf{F}(x) = \left(
\begin{array}{cccc}
a_{1}e^{iq_{1}x}&a_{2}e^{iq_{2}x}&a_{1}e^{-iq_{1}x}&a_{2}e^{-iq_{2}x}\\
b_{1}^{+}e^{iq_{1}x}&b_{2}^{+}e^{iq_{2}x}&b_{1}^{-}e^{-iq_{1}x}&b_{2}^{-}e^{-iq_{2}x}\\
c_{1}e^{iq_{1}x}&c_{2}e^{iq_{2}x}&c_{1}e^{-iq_{1}x}&c_{2}e^{-iq_{2}x}\\
d_{1}^{+}e^{iq_{1}x}&d_{2}^{+}e^{iq_{2}x}&d_{1}^{-}e^{-iq_{1}x}&d_{2}^{-}e^{-iq_{2}x}
\end{array}\right)\cdot
\left(
\begin{array}{c}
\alpha_{1}^{+}\\
\alpha_{2}^{+}\\
\alpha_{1}^{-}\\
\alpha_{2}^{-}
\end{array}\right)
\end{equation}

\noindent or equivalently by matrix blocks 

\begin{equation}\label{eq7}
\mathbf{F}(x)=\left(
\begin{array}{c}
\mathbf{F}_{u}(x)\\
\mathbf{F}_{d}(x)
\end{array}\right)
=\left(
\begin{array}{cc}
\mathbf{U}^{+}(x) & \mathbf{U}^{-}(x)\\
\mathbf{D}^{+}(x) & \mathbf{D}^{-}(x)
\end{array}\right)\cdot
\left(
\begin{array}{c}
\alpha^{+}\\
\alpha^{-}
\end{array}\right),
\end{equation}

\noindent where $\mathbf{F}_{u}(x)$, $\mathbf{F}_{d}(x)$, $\alpha^{+}$ and $\alpha^{-}$ are a two-dimensional column vectors. $\mathbf{U}^{\pm}(x)$ and $\mathbf{D}^{\pm}(x)$ are the $2\times 2$ respective matrix blocks of the $4\times 4$ matrix in equation (\ref{eq6}). We can define the hybrid matrix as 

\begin{equation}\label{eq8}
\left(
\begin{array}{c}
\mathbf{F}_{u}(x_{L})\\
\mathbf{F}_{d}(x_{R})
\end{array}\right)
=\mathbf{H}(x_{R},x_{L})\cdot
\left(
\begin{array}{c}
\mathbf{F}_{d}(x_{L})\\
\mathbf{F}_{u}(x_{R})
\end{array}\right).
\end{equation}

\noindent This equation relates the vectors $\mathbf{F}_{u}(x)$ and $\mathbf{F}_{d}(x)$ at the ends $x_{L}$ and $x_{R}$ of the heterostructure. Note that in the usual hybrid matrix method, we deal with Sturm-Liouville equations with second order derivatives. Here, we have reformulated the procedure to treat first-order equations while maintaining the spirit of the method. Then from equations (\ref{eq7}) and (\ref{eq8}), the hybrid matrix can be written as

\begin{eqnarray}\label{eq12}
\mathbf{H}(x_{R},x_{L}) = \left(
\begin{array}{cccc}
a_{1}e^{iq_{1}x_{L}}&a_{2}e^{iq_{2}x_{L}}&a_{1}e^{-iq_{1}x_{L}}&a_{2}e^{-iq_{2}x_{L}}\\
b_{1}^{+}e^{iq_{1}x_{L}}&b_{2}^{+}e^{iq_{2}x_{L}}&b_{1}^{-}e^{-iq_{1}x_{L}}&b_{2}^{-}e^{-iq_{2}x_{L}}\\
c_{1}e^{iq_{1}(x_{R})}&c_{2}e^{iq_{2}(x_{R})}&c_{1}e^{-iq_{1}(x_{R})}&c_{2}e^{-iq_{2}(x_{R})}\\
d_{1}^{+}e^{iq_{1}(x_{R})}&d_{2}^{+}e^{iq_{2}(x_{R})}&d_{1}^{-}e^{-iq_{1}(x_{R})}&d_{2}^{-}e^{-iq_{2}(x_{R})}
\end{array}\right)\\
\cdot\left(
\begin{array}{cccc}
\nonumber
c_{1}e^{iq_{1}x_{L}}&c_{2}e^{iq_{2}x_{L}}&c_{1}e^{-iq_{1}x_{L}}&c_{2}e^{-iq_{2}x_{L}}\\
d_{1}^{+}e^{iq_{1}x_{L}}&d_{2}^{+}e^{iq_{2}x_{L}}&d_{1}^{-}e^{-iq_{1}x_{L}}&d_{2}^{-}e^{-iq_{2}x_{L}}\\
a_{1}e^{iq_{1}(x_{R})}&a_{2}e^{iq_{2}(x_{R})}&a_{1}e^{-iq_{1}(x_{R})}&a_{2}e^{-iq_{2}(x_{R})}\\
b_{1}^{+}e^{iq_{1}(x_{R})}&b_{2}^{+}e^{iq_{2}(x_{R})}&b_{1}^{-}e^{-iq_{1}(x_{R})}&b_{2}^{-}e^{-iq_{2}(x_{R})}
\end{array}\right)^{-1}.
\end{eqnarray}

Eq. (\ref{eq12}) is the hybrid matrix for a homogeneous domain, in our case well or barrier. To obtain the hybrid matrix for a heterostructure it is essential to know the composition rule, for more details see \cite{39,41,46}. Then, in order to know the transmission and transport properties of a heterostructure it is necessary to know the vectors $\mathbf{F}_{u}(x)$ and $\mathbf{F}_{d}(x)$ of equation (\ref{eq7}) at the ends of the heterostructure and the total hybrid matrix. \\

We assume that a wave $\mathbf{F}^{+}_{01}e^{iq_{1}x}$ traveling from the left side hits the left end of the barrier structure and results in reflections $\mathbf{F}^{-}_{01}e^{-iq_{1}x}$ and $\mathbf{F}^{-}_{02}e^{-iq_{2}x}$ in that domain, while at the right end of the barrier structure we have only transmitted waves $\mathbf{F}^{+}_{01}e^{iq_{1}x}$ and $\mathbf{F}^{+}_{02}e^{iq_{2}x}$, then equation (\ref{eq8}) takes the form

\begin{equation}\label{eq15}
\mathbf{M}_{1}+\mathbf{M}_{2}\cdot\left(
\begin{array}{c}
r_{1}\\
r_{2}\\
t_{1}\\
t_{2}
\end{array}\right)=
\mathbf{H}(x_{R},x_{L})\cdot\mathbf{M}_{3}+\mathbf{H}(x_{R},x_{L})\cdot\mathbf{M}_{4}\cdot\left(
\begin{array}{c}
r_{1}\\
r_{2}\\
t_{1}\\
t_{2}
\end{array}\right),
\end{equation}

\noindent where $r_{1}, r_{2}, t_{1}$ and $t_{2}$ are defined in terms of the coefficients $\alpha^{+}$ and $\alpha^{-}$ at each end.\cite{Briones-Torres2020} In addition, the matrices $\mathbf{M}_{1},\mathbf{M}_{2}, \mathbf{M}_{3}$ and $\mathbf{M}_{4}$ are given as 

\begin{equation}\label{eq16}
\mathbf{M}_{1}=\left(
\begin{array}{c}
a_{1L}\\
b^{+}_{1L}\\
0\\
0
\end{array}
\right),\quad
\mathbf{M}_{2}=\left(
\begin{array}{cccc}
a_{1L}&a_{2L}&0&0\\
b^{-}_{1L}&b^{-}_{2L}&0&0\\
0&0&c_{1R}&c_{2R}\\
0&0&d^{+}_{1R}&c^{+}_{2R}
\end{array}
\right),
\end{equation}

\begin{equation}\label{eq17}
\mathbf{M}_{3}=\left(
\begin{array}{c}
c_{1L}\\
d^{+}_{1L}\\
0\\
0
\end{array}
\right);\quad
\mathbf{M}_{4}=\left(
\begin{array}{cccc}
c_{1L}&c_{2L}&0&0\\
d^{-}_{1L}&d^{-}_{2L}&0&0\\
0&0&a_{1R}&a_{2R}\\
0&0&b^{+}_{1R}&b^{+}_{2R}
\end{array}
\right).
\end{equation}

\noindent The subscripts $L$ and $R$ indicate the external domain where the components of the wavefunction amplitudes are calculated. Finally, 

\begin{equation}\label{eq18}
\left(
\begin{array}{c}
r_{1}\\
r_{2}\\
t_{1}\\
t_{2}
\end{array}\right)=
\left[\mathbf{M}_{2}-\mathbf{H}(x_{R},x_{L})\cdot\mathbf{M}_{4}\right]^{-1}
\cdot\left[\mathbf{H}(x_{R},x_{L})\cdot\mathbf{M}_{3}-\mathbf{M}_{1}\right].
\end{equation}

\noindent Therefore, the transmittance is given by

\begin{equation}
\mathbb{T}(E,\theta)=\left|t_{1}\right|^2,
\end{equation}

\noindent where $E$ and $\theta$ are the energy and angle of incidence of the electrons in the bilayer graphene structure.\\ 

To calculate the transport and thermoelectric properties it is necessary to obtain the conductance, this can be done using the Landauer-B\"uttiker formalism.\cite{48} Under this formalism, the linear-regime conductance is obtained by summing over all transmission channels

\begin{equation}
\frac{G}{G_{0}}=E_{F}\int_{-\frac{\pi}{2}}^{\frac{\pi}{2}}\mathbb{T}(E_{F},\theta)\cos{\theta} d\theta,
\label{conductance1}
\end{equation}

\noindent where $E_{F}$ is Fermi energy and $G_{0}=\frac{2e^{2}L_{y}}{h^{2}v_{F}}$ the  fundamental conductance factor with $e$, $L_{y}$ and $h$ as the electron charge, the width of bilayer graphene sheet in the transverse y-coordinate. For our calculations, the width of the bilayer graphene is set to $L_{y} = 200$ nm.\cite{22}\\

In the case of thermoelectric properties, it is necessary to know the Seebeck coefficient, it can be done with the help of Cutler-Mott equation,\cite{18,22,49,50} 

\begin{equation}
S(E,\theta)=\left. \frac{\pi^{2}k_{B}^{2}T}{3e}\frac{\partial \ln G(E)}{\partial E}\right|_{E=E_{F}},
\label{cuttler}
\end{equation}

\noindent where $e$ is the electron charge, $k_{B}$ is the Bolzmann constant, $T$ is the average temperature between the hot and cold sides of the bilayer graphene structure, in our case we take $T=50$ K.\\

Finally, for a material to be a good thermoelectric, its power factor needs to be high. The power factor is defined by the squared magnitude of the Seebeck coefficient and the conductance, $S^{2}G$.\cite{18,22}

\section*{Results and Discussion}

Recent studies in bilayer graphene superlattices have shown that Fano and hybrid resonances leave an identifiable characteristic mark on transport properties.\cite{39} It was also shown that the opening of the bandgap enhances the Fano-resonance response of bilayer graphene single and double barrier structures.\cite{Briones-Torres2020} In view of this, it is essential to analyze whether thermoelectric properties improve or perhaps exceed current reported values. The thermoelectric properties in our case are calculated using Cutler-Mott formula, which implies the derivative of a logarithmic function, this means that the asymmetric profile of the Fano or hybrid resonances could make bilayer graphene an excellent thermoelectric material. 

Under this context, we firstly show the results for the Seebeck coefficient and the power factor of bilayer graphene single barriers. Secondly, we analyze the corresponding results of bilayer graphene double barriers. We paid special attention to the impact of Fano and hybrid resonances on the mentioned thermoelectric properties. We also compare the thermoelectric properties of gapless and gapped bilayer graphene barrier structures. 

\subsection*{Bilayer graphene single barriers}

\begin{figure}[htb]
\centering
\includegraphics[scale=0.6]{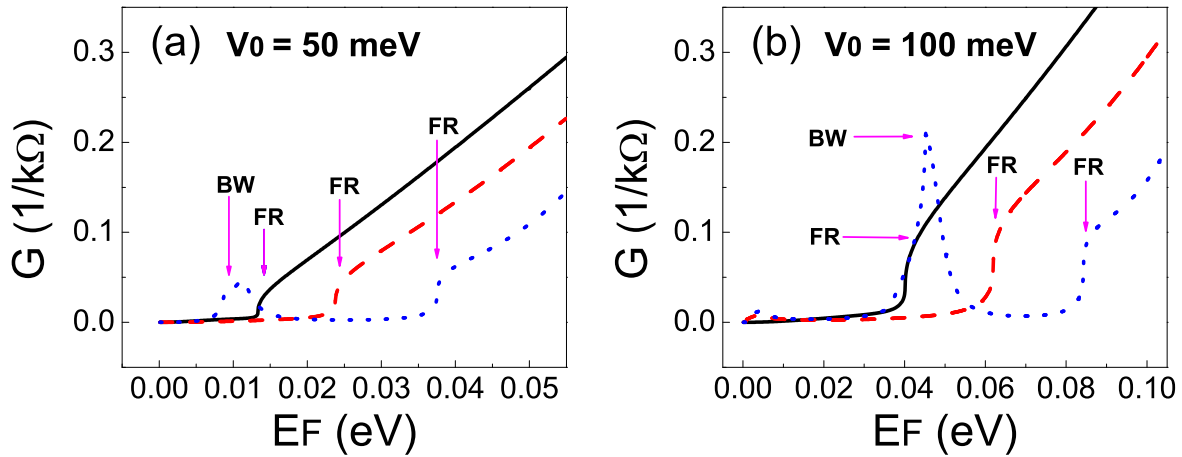}
\caption{\label{Fig1} Conductance versus the Fermi energy for gapless bilayer graphene single barriers. Two different heights of the barrier are considered: (a) 50 meV and (b) 100 meV. As there in no bandgap, $V_1=V_2=V_0$. The solid-black, dashed-red and dotted-blue curves correspond to barrier widths of 6 nm, 10 nm and 20 nm, respectively. The arrows refer to the type of resonance that predominantly contribute to the conductance, being FR and BW the abbreviations for Fano and Breit-Wigner resonances.}
\end{figure}

At first place, we consider the transport and thermoelectric properties of gapless single barriers, that is, the potentials are equal $V_{1}=V_{2}=V_{0}$, and consequently there is no bandgap opening. We will begin by analyzing the energy regions where Fano resonances occur in electron transport. Figure \ref{Fig1} shows the conductance as a function of Fermi energy for different barrier widths. Two typical heights of the potential barrier are considered: (a) 50 meV and (b) 100 meV. The pink arrows indicate the energy regions where the resonances of the system leave a mark on the conductance. FR and BW stand for Fano and Breit-Wigner resonances, respectively. As we can see, for Breit-Wigner resonances wide conductance peaks take place, while for Fano resonances a sudden rise in the conductance arise regardless of the height of the barrier.\cite{39} The fundamental modifications when changing from $V_{0}=50$ meV to $V_{0}=$100 meV are: the regions associated to Fano and Breit-Wigner resonances shift to higher energies, the sudden rise associated to Fano resonances is softened and the peak related to Breit-Wigner resonances is more evident. 

\begin{figure}[htb]
\centering
\includegraphics[scale=0.6]{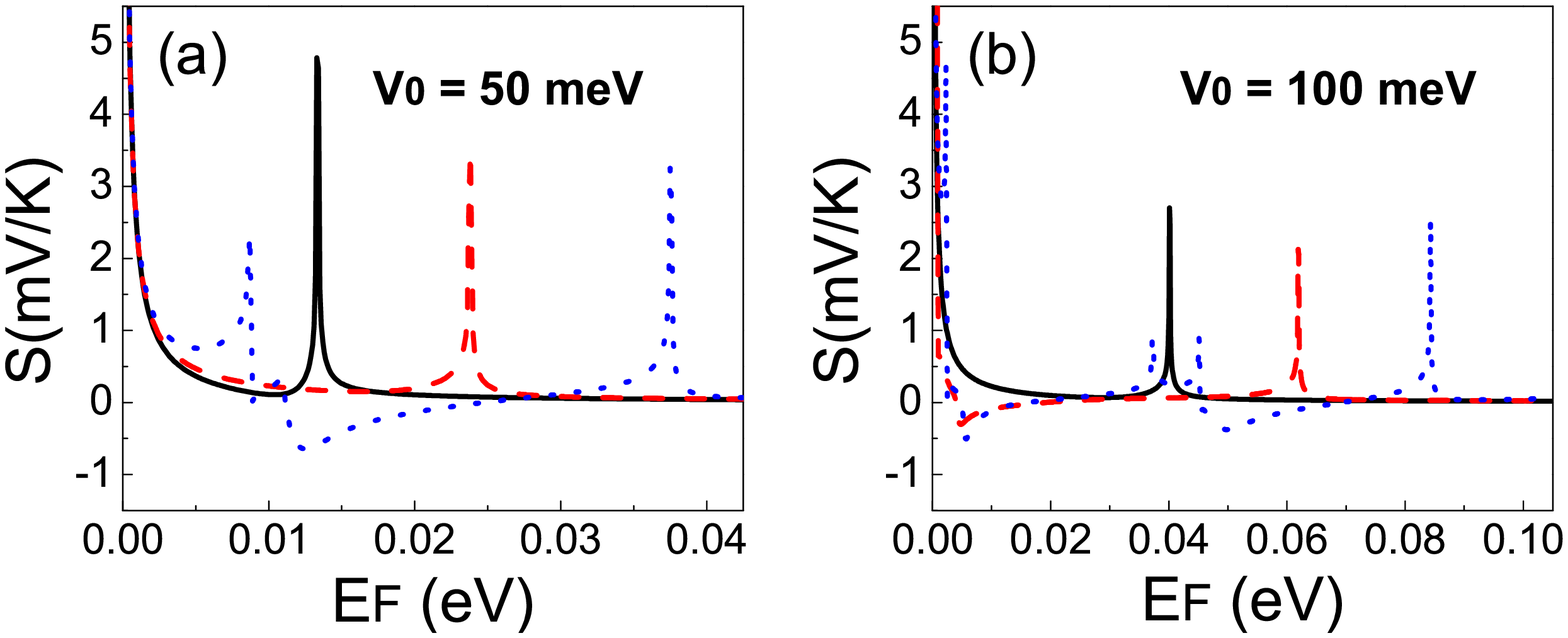}
\caption{\label{Fig2} The Seebeck coefficient versus the Fermi energy for gapless bilayer graphene single barriers. The system parameters are the same as in Fig. \ref{Fig1}.}
\end{figure}

The contribution of the Fano resonances to the thermoelectric properties, particularly to the Seebeck coefficient, is shown in figure \ref{Fig2}. The heights and widths of the barriers correspond to the ones considered in Fig. \ref{Fig1}. As we can see the Seebeck coefficient is maximized in the energy regions in which the Fano resonances are preponderant. So, in correspondence with the conductance curves, the peak of the Seebeck coefficient shifts to higher energies as both the barrier width and the barrier height increases. In addition, regardless of the height of the barrier, the Seebeck coefficient presents a decreasing trend as the barrier width gets larger. Likewise, all Seebeck peaks diminish when we rise the height of the barrier. This diminshment of the Seebeck coefficient obeys the deformation of the asymmetric profile of Fano resonance as both the barrier width and barrier height grows.\cite{39} Regarding Breit-Wigner resonances, we can see that its contribution to the Seebeck coefficient is not as prominent as the one of Fano resonances. For instance, although BW resonances contribute significantly to the conductance of a barrier of 100 meV and 20 nm, see the huge peak around 45 meV in the dotted-blue curve of Fig. \ref{Fig1}b, the Seebeck coefficient associated to these resonances is not as important as the peak related to Fano resonances, compare the Seebeck coefficient peaks around 45 meV and 85 meV in the dotted-blue curve of Fig. \ref{Fig2}b. This remarkable difference of the impact of FR and BW resonances in the thermoelectric properties of bilayer graphene single barriers is a consequence of the rate of change of the linear-regime conductance. In fact, the rate of change of the conductance is larger in FR energy regions than in BW ones. It is also important to remark the magnitude of the Seebeck coefficient associated to Fano resonances. As we can notice in the vertical scale of Fig. \ref{Fig2}, in practically all single barriers studied, the Seebeck coefficient reaches values of mV/K. These values are of the order of the ones reported for gated silicene superlattices\cite{22} and three orders of magnitude larger than the ones for gated graphene superlattices.\cite{18}

\begin{figure}[htb]
\centering
\includegraphics[scale=0.6]{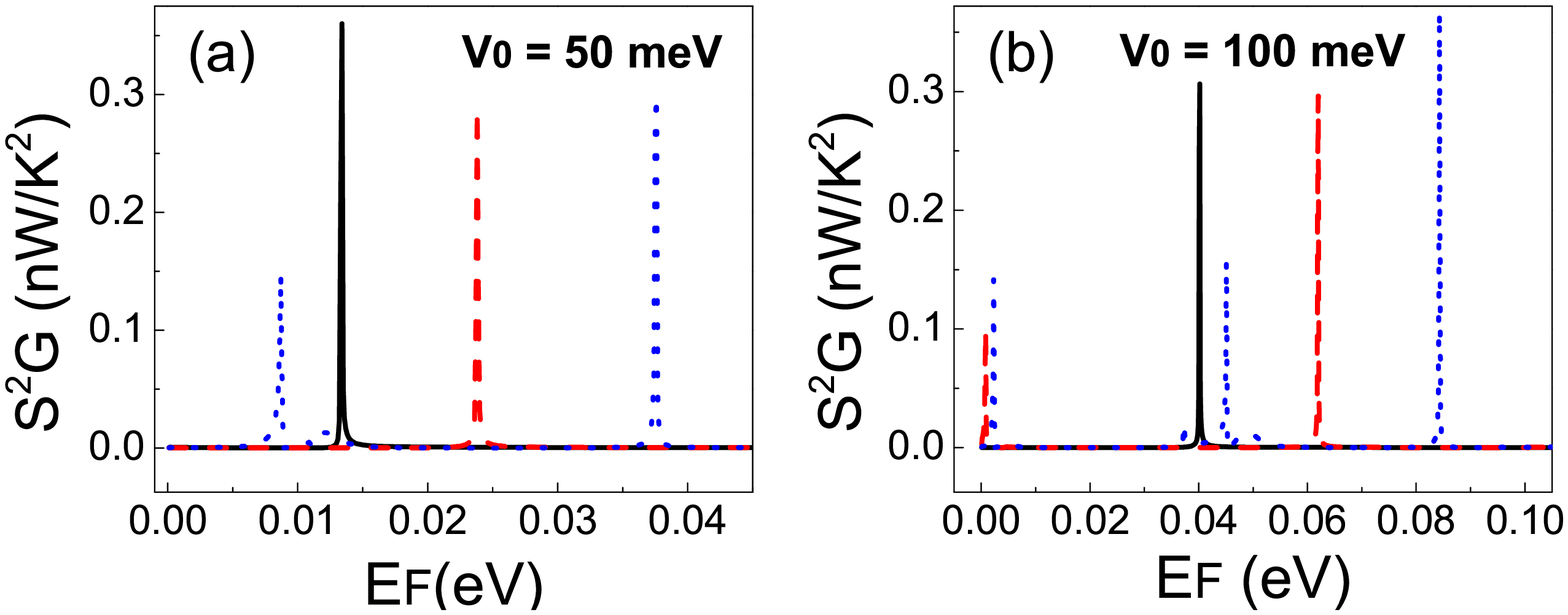}
\caption{\label{Fig3} Power factor versus the Fermi energy for gapless bilayer graphene single barriers. The system parameters are the same as in Fig. \ref{Fig1}.}
\end{figure}

With the calculation of the conductance and the Seebeck coefficient we can obtain straightforwardly the power factor $S^2G$. In fact, the corresponding results of the power factor are shown in Fig. \ref{Fig3}. As we can see the power factor is also maximized in the energy regions in which the Fano resonances are preponderant. However, in this case the trend of the power factor with the height and width of the barriers is not as simple as in the case of the Seebeck coefficient. The power factor is the result of the interplay between the square of the Seebeck coefficient and the conductance. So, it is not sufficient to have a high Seebeck coefficient to obtain a high power factor, likewise, it is not enough to have a large conductance to obtain a high power factor. Actually, it is the combination of both, the Seebeck and the conductance, that shapes the power factor. For instance, in the case of single barriers of a height of 50 meV the highest power factor is presented for a width of 6 nm, see the solid-black curve in Fig. \ref{Fig3}a. However, there is no a systematic reduction of the power factor as the barrier width increases. Actually, the power factor for barriers of 20 nm is a bit greater than the one of barriers of 10 nm. Furthermore, for barriers of 100 meV the dynamic is reverse, that is, the highest power factor is presented for a width of 20 nm, while the power factor for barriers of 6 nm is a bit greater than the one for barriers of 10 nm, compare the peaks in Fig. \ref{Fig3}b. We can also see that regardless of the barrier height the power factor peaks associated to BW resonances are half the value of the peaks related to Fano resonances. With regard of the magnitude of the power factor, the maximum value 0.35 nW/K$^2$ found for single barriers represents twice the best values reported in gated graphene superlattices,\cite{18} however it is far from the optimum values for gated silicene superlattices.\cite{22}

\begin{figure}[htb]
\centering
\includegraphics[scale=0.65]{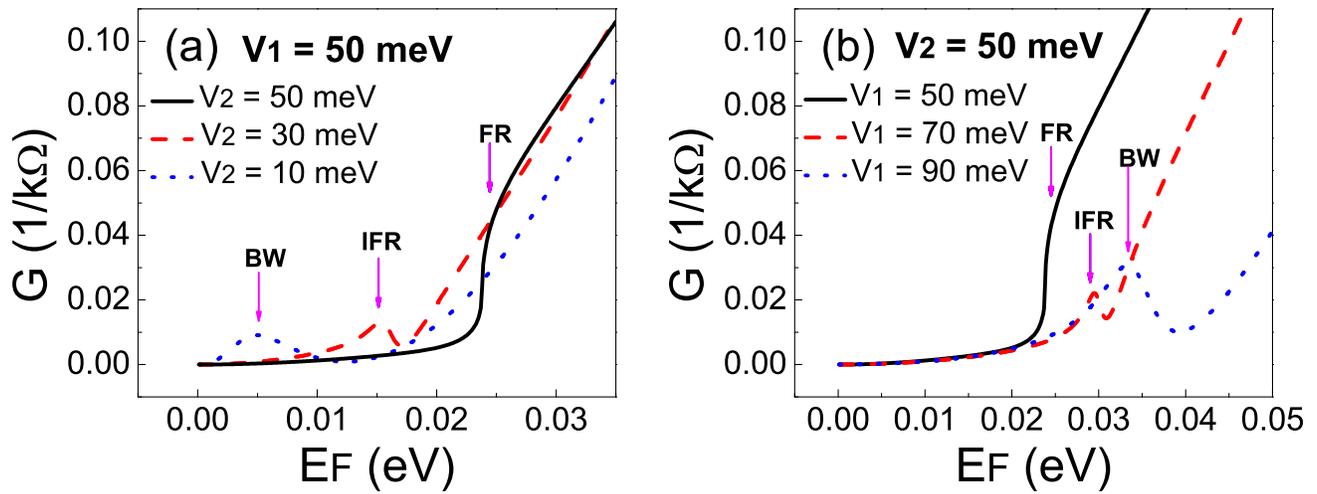}
\caption{\label{Fig4} Conductance versus the Fermi energy for different values of the bandgap. In (a) we have fixed $V_{1}$ and varied $V_{2}$, while in (b) $V_{2}$ is fixed and $V_{1}$ is varied. The width of the barrier is 10 nm. The arrows indicate the type of resonance, BW for Breit-Wigner, FR for Fano resonance and IFR inverted Fano resonance.}
\end{figure}

\begin{figure}[htb]
\centering
\includegraphics[scale=0.35]{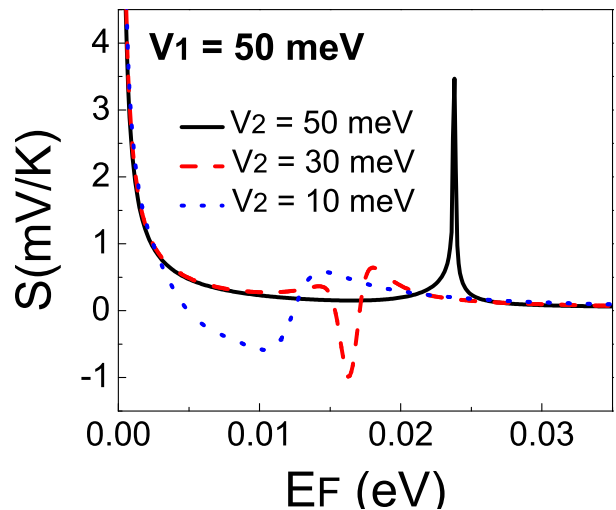}
\caption{\label{Fig5} Seebeck coefficient versus the Fermi energy for different values of the bandgap. The system parameters are the same as in Fig. \ref{Fig4}a.}
\end{figure}

Regarding gapped bilayer graphene single barriers, it is reported that the bandgap opening activates additional transmission channels with Fano characteristics, resulting in an effective enhancement of the Fano-resonance response of the transport properties.\cite{Briones-Torres2020} Specifically, the hallmark of Fano resonances is directly manifested in the conductance. The fundamental changes caused by the bandgap opening are shown in Fig. \ref{Fig4}. As we can notice once the bandgap is opened the conductance curve of gapless barriers is deformed, giving place to a Fano-like profile (IFR) on the conductance. In fact, the conductance Fano-like profile is shifted to lower energies if $V_1$ is fixed and $V_2$ is varied, while it is shifted to higher energies if $V_2$ is fixed and $V_1$ is varied, see Fig. \ref{Fig4}a and b, respectively. So, the bandgap opening can be used as a modulation parameter to corroborate the exotic phenomenon of Fano resonances in bilayer graphene single barriers. However, it is not clear at all if the bandgap opening can help to improve the thermoelectric properties of bilayer graphene single barriers, so, we proceed to analyze it. As the effect on the transport properties is practically the same when $V_{1}$ or $V_{2}$ varies, we will only consider one of them to analyze the Seebeck coefficient and the power factor. In Fig. \ref{Fig5} we show the Seebeck coefficient results that correspond to the transport properties of Fig. \ref{Fig4}a. As we can see the Seebeck coefficient of gapless single barriers is greater than the corresponding one to gapped barriers. The reduction of the Seebeck coefficient obeys the systematic deformation of the gapless conductance curve as the bandgap increases. In fact, the Seebeck coefficient responds to the logarithmic derivative of the conductance, thereby, the transformation of the abrupt conductance profile of gapless barriers into a smooth Fano-like profile for gapped barriers results in a significant reduction of the Seebeck coefficient. Specifically, the Seebeck passes from 4 mV/K for gapless barriers to less than 1 mV/K for gapped ones.

\begin{figure}[htb]
\centering
\includegraphics[scale=0.6]{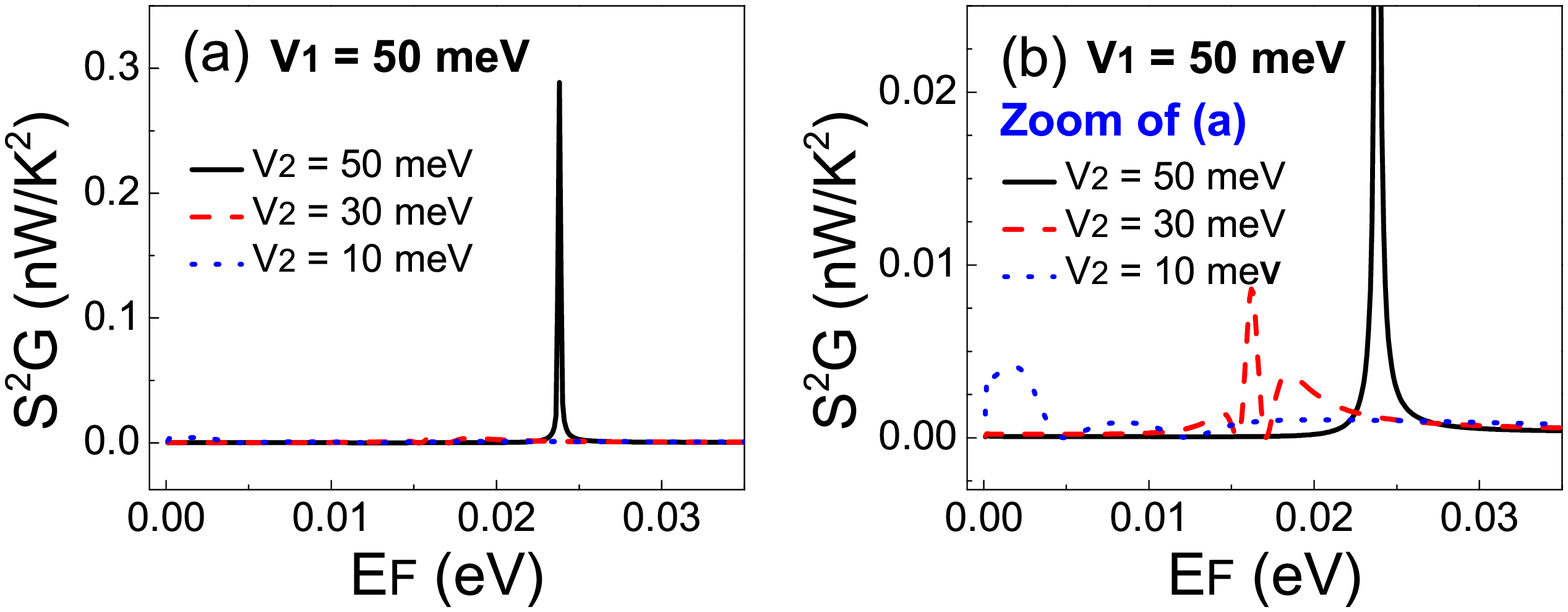}
\caption{\label{Fig6} Power factor versus the Fermi energy for different values of the bandgap. The system parameters are the same as in Fig. \ref{Fig4}a.}
\end{figure}

In the case of power factor, we find a similar behavior with the bandgap opening as for the Seebeck coefficient. In Fig. \ref{Fig6}a the power factor results for different bandgap openings are shown. The power factor curves correspond to the conductance and Seebeck coefficient outputs of Figs. \ref{Fig4} and \ref{Fig5}, respectively. As we can notice the power factor of gapless barriers is dominant with a value of roughly 0.3 nW/K$^2$. It is worth noting that the power factor curves for gapped barriers are barely noticeable. If we amplify the power factor curves of gapped barriers, we can realize that the bandgap opening reduces the power factor values more than an order of magnitude, values less than 0.01  nW/K$^2$. So, as we have documented, the bandgap opening is not beneficial for the thermoelectric properties of bilayer graphene single barriers and consequently a precise control of the potentials in the graphene layers is required to preserve the significant thermoelectric properties of gapless barriers. 

\subsection*{Bilayer graphene double barriers}

\begin{figure}[htb]
\centering
\includegraphics[scale=0.6]{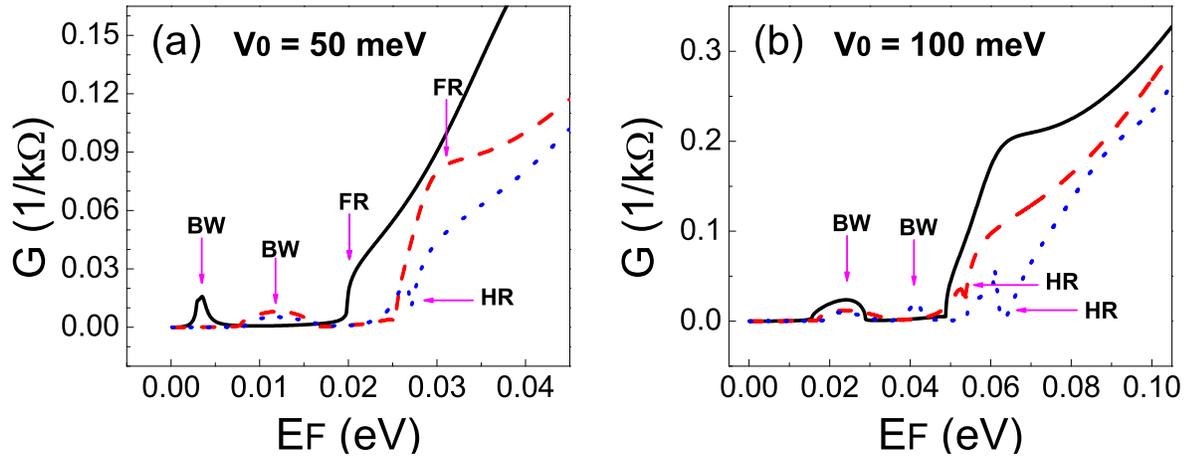}
\caption{\label{Fig7} Conductance versus the Fermi energy for gapless bilayer graphene double barriers. Two different heights of the barrier are considered: (a) 50 meV and (b) 100 meV. As there in no bandgap, $V_1=V_2=V_0$. In the case of barriers of 50 meV, the solid-black, dashed-red and dotted-blue curves correspond to barrier-well widths of 6 nm, 9 nm and 10 nm, respectively, while for barriers of 100 meV the same type of curves correspond to 6 nm, 7 nm and 9 nm. The arrows refer to the type of resonance that predominantly contribute to the conductance: FR, BW and HR stand for Fano, Breit-Wigner and Hybrid resonances, respectively.}
\end{figure}

Now, it is time to analyze the case of gapless double barriers. Throughout this subsection we will deal with symmetric barriers, that is, the height of the barriers is the same as well as the widths of the barriers and the well. It is worth mentioning that in this case the interplay between Fano resonances and resonant states of the quantum well region give rise to the so-called hybrid resonances (HR).\cite{39} Depending on the structural parameters of the double barriers the transport properties can present energy regions in which BW, FR and HR contribute predominantly. In Fig. \ref{Fig7} we show the conductance as a function of Fermi energy for different widths of barriers-well. We have considered the same barrier heights as in the case of single barriers: (a) 50 meV and (b) 100 meV. As we can see the mentioned resonances are manifested in the conductance. BW and FR resonances result, as in the case of single barriers, in wide peaks  and a sudden rise in the conductance, while HR resonances give rise to a peak followed by a minimum. In particular, for barriers of 50 meV and widths of 6 nm and 9 nm, BW and FR resonances dominate the conductance characteristics, see the solid-black and dashed-red curves in Fig. \ref{Fig7}a. In the case of barriers of 10 nm, BW and HR resonances shape the conductance curve profile, see see the dotted-blue curve in Fig. \ref{Fig7}a. If we increase the height of the barriers, the coupling between Fano resonances and the resonant states of the well region is more effective and consequently the signatures of HR more evident on the transport properties. For instance, we can clearly see the energy region associated to HR for double barriers of 100 meV and widths of 7 nm and 9 nm, dashed-red and dotted-blue curves in Fig. \ref{Fig7}b, respectively. In the case of double barriers of 6 nm, the sudden rise no longer fully represents the contribution of FR, actually, represents a transition between FR and HR. We are not considering the case of 10 nm because the signatures of HR resonances are not longer present. For more details about the transport characteristics of gapless double barriers see Ref. 39. 

\begin{figure}[htb]
\centering
\includegraphics[scale=0.6]{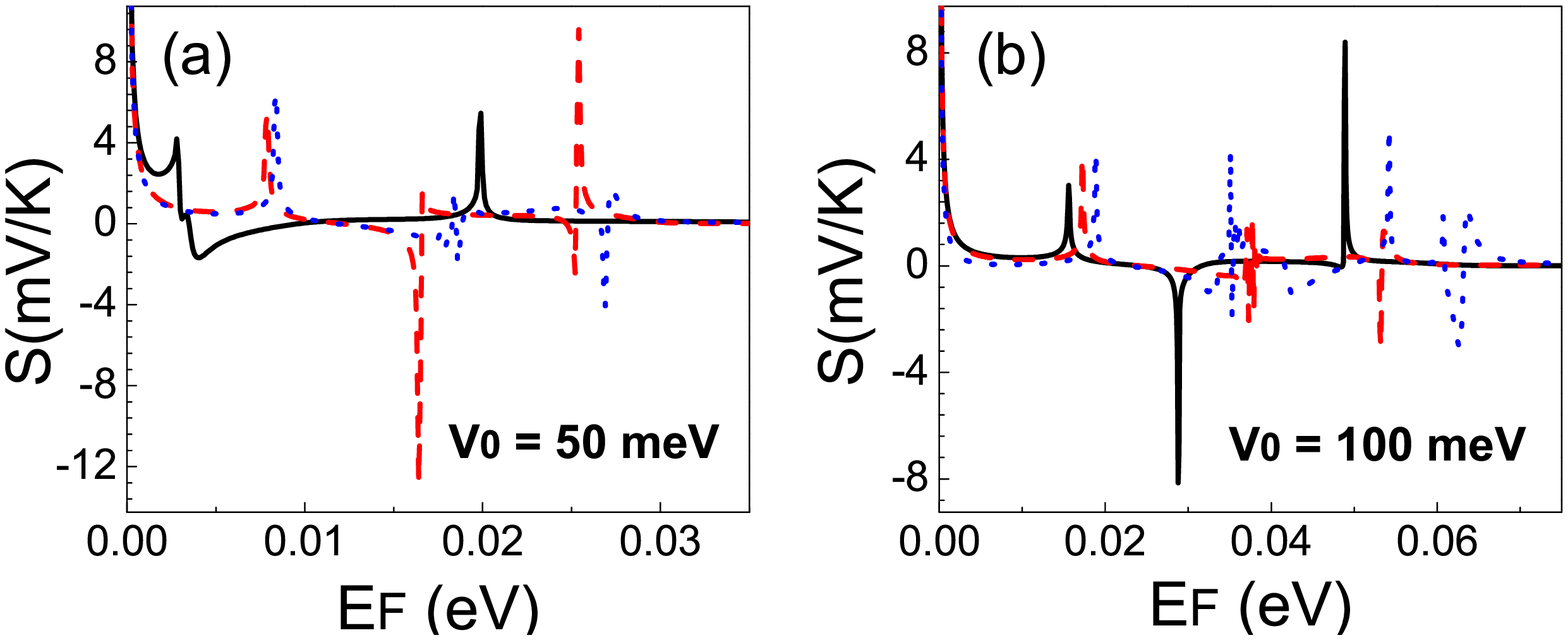}
\caption{\label{Fig8} Seebeck coefficient versus the Fermi energy for gapless bilayer graphene double barriers. The system parameters are the same as in Fig. \ref{Fig7}.}
\end{figure}

After analyzing the fundamental characteristics of the conductance curves of gapless double barriers, we proceed to present the results of the thermoelectric properties. In first place,  in Figure \ref{Fig8} we show the Seebeck coefficient results. The parameters are the same as in Figure \ref{Fig7}. Like the case of single barriers, there is a correspondence between the conductance characteristics and the Seebeck coefficient ones. In fact, the Seebeck coefficient peaks shift to higher energies as the height of the barriers and the width of barriers-well increase. We can also notice that  the contribution of HR resonances to the Seebeck coefficient is not superior to the contribution associated to FR resonances, or even to the one related to BW resonances. For instance, HR resonances give rise to a Seebeck coefficient of about 3 mV/K for barriers of 50 meV and 10 nm, see the peak about 26 meV in the blue-dotted curve of Fig. \ref{Fig8}a. This value is 3 and 4 times smaller than the values related to FR and BW resonances respectively, see the peaks close to 16 meV and 25 meV in the dashed-red curve of Fig. \ref{Fig8}a. Even, this value is smaller than the best Seebeck coefficient values found for single barriers, compare with Fig. \ref{Fig2}. Here, it is also worth mentioning that the Seebeck coefficient peaks associated to FR and BW resonances are 2 and 3 times larger than the corresponding ones to single barriers. This increase obeys the better definition of the conductance curves of bilayer graphene double barriers. In particular, to the more abrupt transition between conductance gaps and the regions in which FR and BW are preponderant, keep in mind that the Seebeck coefficient is the logarithmic derivative of the conductance. For barriers of 100 meV the same dynamic is presented, being FR and BW resonances dominant. In fact, we can see that the contribution of FR and BW resonances for widths of 6 nm reaches values of 8 mV/K, see the solid-black curve in Fig. \ref{Fig8}. Furthermore, in spite HR resonances dominate the transport characteristics of double barriers with 9 nm its Seebeck coefficient is not as large as the one for barriers of 6 nm. It is also important to mention that, as in the case of single barriers, the Seebeck coefficient peaks diminish as the height of the barriers increases. 

\begin{figure}[htb]
\centering
\includegraphics[scale=0.6]{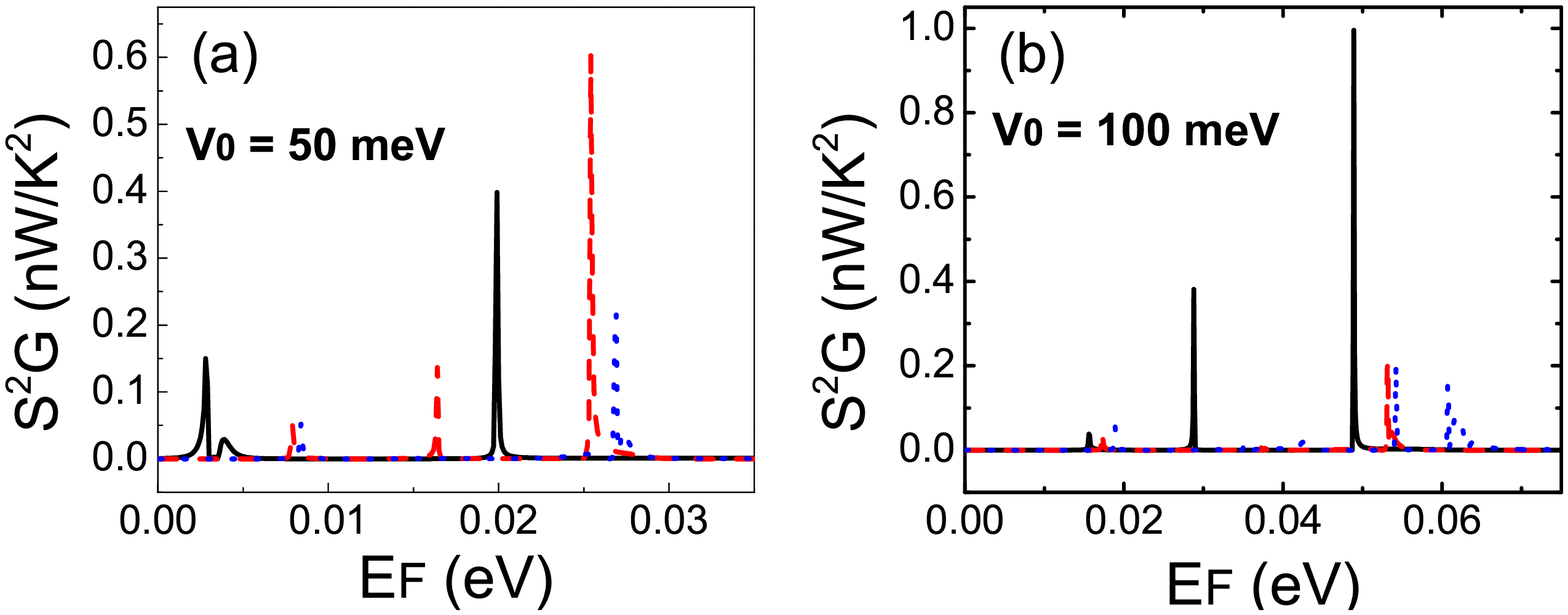}
\caption{\label{Fig9} Power factor versus the Fermi energy for gapless bilayer graphene single barriers. The system parameters are the same as in Fig. \ref{Fig7}.}
\end{figure}

To close the results part for gapless double barriers we analyze the main characteristics of the power factor. In Fig. \ref{Fig9} the power factor results that correspond to the conductance and Seebeck coefficient of Figs. \ref{Fig7} and \ref{Fig8} are shown. As we can notice there is a direct correspondence between the energy regions with maximum power factor and the energy regions associated to the different resonances that gapless double barriers harbor. In fact, the power factor characteristics are dominated by FR and BW resonances, as in the case of the Seebeck coefficient. However, as the power factor is the product of the square of the Seebeck coefficient and the conductance turns out that FR resonances are clearly dominant over BW resonances. For instance, in the case of barriers of 50 meV, we can notice that the power factor peak associated to FR resonances is 3 and 5 times larger than the one of BW resonances for widths of 6 nm and 9 nm, respectively. We can also see that in this case the power factor peak related to HR resonances is a bit greater than the peaks of BW resonances. This is quite interesting because the Seebeck coefficient of BW resonances is significantly larger than the one of HR resonances, compare the dashed-red and dotted-blue curves in Fig. \ref{Fig8}a. So, what really matters for the power factor is the product of the Seebeck coefficient and the conductance, hence, a large Seebeck coefficient is a good signal, but not enough to guarantee a high power factor. We can notice something similar for barriers of 100 meV. For instance, the Seebeck coefficient peaks for barriers of 6 nm are not as prominent as the peaks for barriers of 50 meV and 9 nm, compare the solid-black curve in Fig. \ref{Fig8}b and the dashed-red curve in Fig. \ref{Fig8}a. However, the power factor peaks of the former are significantly greater than the corresponding ones of the latter. Actually, the power factor peak associated to FR resonances for barriers of 100 meV and 6 nm is the highest one found in the present work, about 1 nW/K$^2$.    

\begin{figure}[htb]
\centering
\includegraphics[scale=0.65]{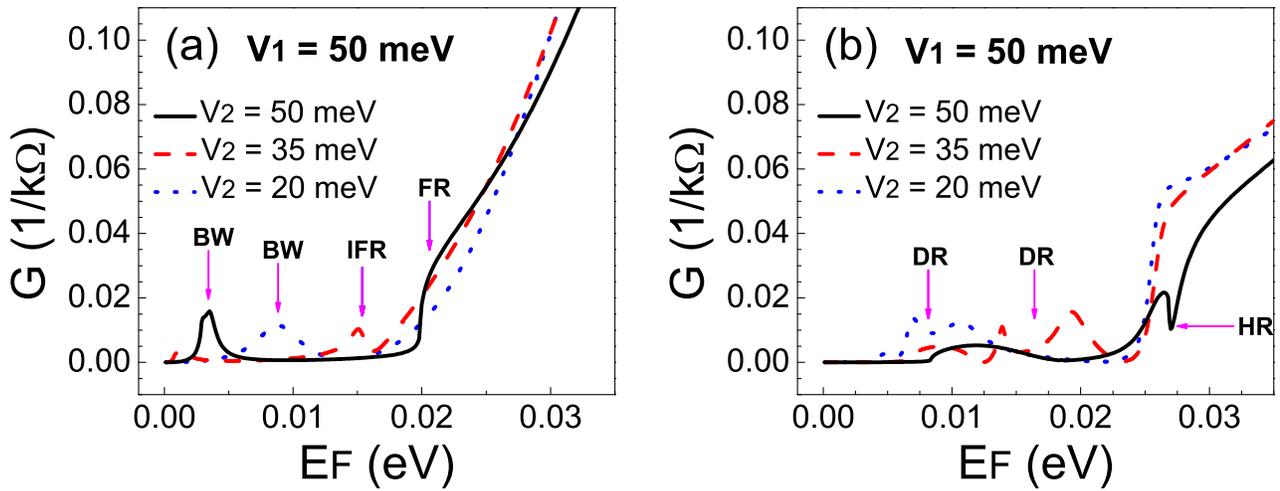}
\caption{\label{Fig10} Conductance versus the Fermi energy for different values of the bandgap. We have fixed $V_{1}$ and varied $V_{2}$. In (a) the width of the barriers and well is 6 nm, while in (b) is 10 nm. The arrows indicate the type of resonances: BW for Breit-Wigner resonances, FR for Fano resonances, IFR inverted Fano resonances, HR for hybrid resonances and DR for double resonances.}
\end{figure}

\begin{figure}[htb]
\centering
\includegraphics[scale=0.65]{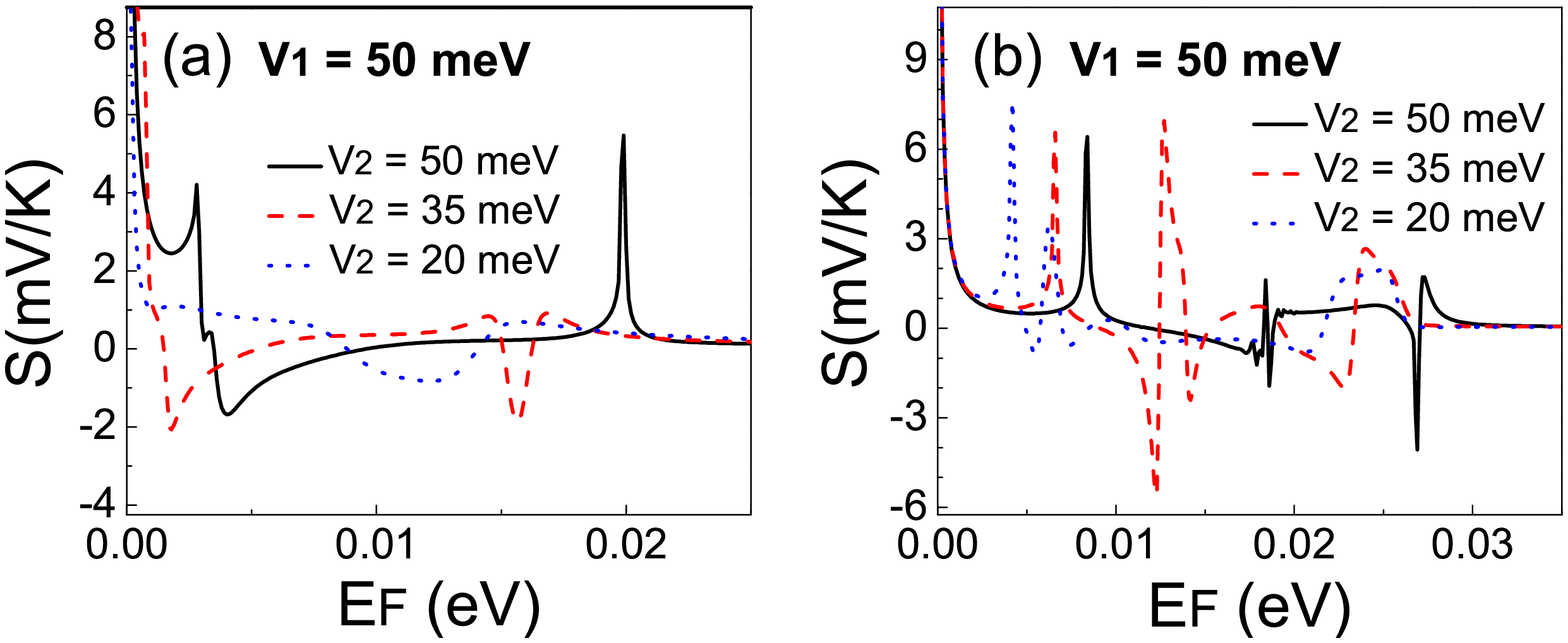}
\caption{\label{Fig11} Seebeck coefficient versus the Fermi energy for different values of the bandgap. In (a) the width of the barriers and well is 6 nm, while in (b) is 10 nm. The system parameters are the same as in Fig. \ref{Fig10}.}
\end{figure}

\begin{figure}[htb]
\centering
\includegraphics[scale=0.6]{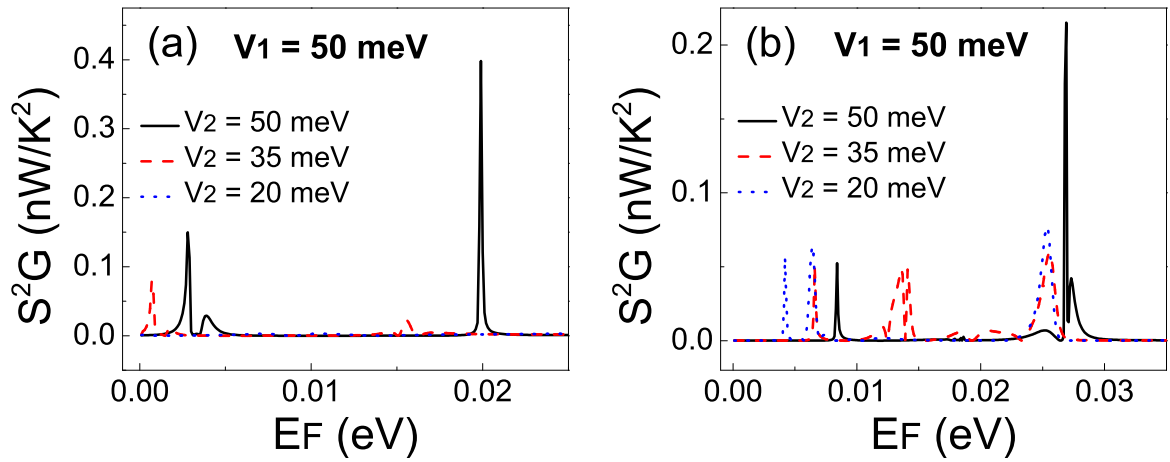}
\caption{\label{Fig12} Power factor versus the Fermi energy for different values of the bandgap. In (a) the width of the barriers and well is 6 nm, while in (b) is 10 nm. The system parameters are the same as in Fig. \ref{Fig10}.}
\end{figure}

To finish, we analyze the bandgap opening effect on the thermoelectric properties of double barriers. As in the case of single barriers, we only take one case for the bandgap opening, that is, $V_{1}$ is fixed to 50 meV and $V_2$ is varied. However, as the dynamics in matter of resonances is interesting, we address two widths of barriers-well, 6 nm and 10 nm, Fig. \ref{Fig10}a and b, respectively. For barriers of 6 nm, the bandgap opening activates additional channels that give rise to a Fano-like profile directly on the conductance that we named (labeled) as inverted Fano resonances (IFR). Depending on the magnitude of the bandgap the main contributions to the transport are due to FR, BW and IFR resonances.\cite{Briones-Torres2020} For barriers of 10 nm, the systematic deformation of HR resonances as well as the interplay with BW resonances result in the so-called double resonances (DR). In fact, the conductance characteristics are dominated by HR and DR resonances.\cite{Briones-Torres2020} The corresponding results for the Seebeck coefficient are shown in Fig. \ref{Fig11}. For barriers of 6 nm, the Seebeck coefficient decreases as the bandgap gets larger. This is due to the systematic deformation of the abrupt profile of the conductance curves related to FR and BW resonances.\cite{Briones-Torres2020} In particular, the peak around 0.015 eV is due to IFR resonances and is similar to the one at 0.02 eV, which is due to FR resonances. Another interesting aspect is that for gapped barriers the Seebeck coefficient peaks associated to BW resonances are not as significant as for gapless barriers. With regard of barriers of 10 nm, the dominant contribution is due to DR resonances, particularly a bandgap aperture of 15 meV. Even, DR resonances perform better than HR resonances. However, as we have insisted throughout the present study, a good Seebeck coefficient not necessary guarantee a significant power factor. The power factor results are shown in Fig. \ref{Fig12}. As we can see the bandgap opening does not improve the thermoelectric properties at all. Actually, the FR resonances of gapless barriers are the ones that considerably maximize the power factor. See the power factor results for 6 nm, solid-black curve in Fig. \ref{Fig12}a. Furthermore, for barriers of 10 nm, the contribution of DR resonances is not as preponderant as the one related to HR resonances, compare the conductance curves in Fig. \ref{Fig12}b. So, as we have documented the thermoelectric properties of gapped double barriers are far from the gapless counterparts. In particular, far from the bests value which is 1.0 nW/K$^2$.

%\subsection*{Discussion and important remarks}

%\begin{enumerate}

%\item 

%\item 

%\item 

%\end{enumerate}
  
\section*{Conclusions}

In summary, we have analyzed the transport and thermoelectric properties of bilayer graphene single and double barrier structures. In particular, we have assessed the impact of Fano and hybrid resonances of gated bilayer graphene barriers on the Seebeck coefficient and power factor. We have described the charge carriers in bilayer graphene as massive chiral particles through a four-band Hamiltonian. We have used a modified version of the hybrid matrix method, the Landauer- B\"uttiker formalism and the Cutler-Mott formula to obtain the transmission, transport and thermoelectric properties, respectively. We have compared the thermoelectric properties of gapless and gapped bilayer graphene barrier structures. We found that the thermoelectric properties are significantly enhanced in the energy regions in which the Fano resonances occur. In particular, the Seebeck coefficient and the power factor reached values of mV/K and nW/K$^2$, which are similar to the values reported for gated silicene superlattices\cite{22} and three orders of magnitude above the ones reported for gated graphene superlattices.\cite{18} Regarding hybrid resonances, the thermoelectric properties are also improved, however the values of the Seebeck coefficient and the power factor are not as prominent as for Fano resonances. In the case of gapped barrier structures, it is reported that the bandgap opening helps to see the signatures of the Fano resonances directly on the transport properties,\cite{Briones-Torres2020} however for the thermoelectric properties the bandgap opening is not beneficial. Specifically, the Seebeck coefficient and power factor of gapped single barriers are systematically deteriorated as the bandgap increases. For gapped double barriers, the bandgap opening modifies the interplay between Fano resonances and resonant states of the well region, giving rise to energy regions with appreciable thermoelectric properties. However, the Seebeck coefficient and power factor values did not surpass the corresponding ones to gapless double barrier structures. We believe that bilayer graphene nanostructures can favor the field of renewable energy and thermoelectricity because bilayer graphene is a stable material and its production method is well established.

\section*{Acknowledgements}

J.A.B.-T. acknowledges to PRODEP-SEP-Mexico for the Posdoctoral Research Fellowship 511-6/2019.-10959. J.A.B.-T. and R.P.-A. are thankful with the Autonomous University of Zacatecas for the hospitality. I.R.-V. would like to acknowledge to CONACYT-SEP Mexico for the financial support through grant A1-S-11655. 

\section*{Author contributions statement}

\textbf{J. A. Briones-Torres:} Methodology, Software, Data curation, Investigation, Writing - Reviewing \& editing. \textbf{R. P\'{e}rez-\'{A}lvarez:} Writing - reviewing \& editing. \textbf{S. Molina-Valdovinos:} Writing - reviewing \& editing. \textbf{I. Rodr\'{i}guez-Vargas:} Conceptualization, Investigation, Writing - original draft, Visualization, Funding acquisition. 

\section*{Additional information}

\textbf{Competing financial interests:} The authors declare no competing financial interest. 

\end{document}